\documentstyle[prb,aps,twocolumn,epsf]{revtex}

\begin{document}
\draft

\flushbottom
\twocolumn[
\hsize\textwidth\columnwidth\hsize\csname @twocolumnfalse\endcsname

\title{Solution of the X-ray edge problem for 2D electrons in a magnetic field}
\author{H. Westfahl Jr.$^1$, A.O. Caldeira$^1$, D. Baeriswyl$^2$, and
E. Miranda$^1$} 
\address{$^1$Instituto de F\'{\i}sica Gleb Wataghin\\
Universidade Estadual de Campinas\\
13083-970, Campinas, SP, Brazil\\
$^2$Institut de Physique Th\'{e}orique\\
Universit\'{e} de Fribourg, P\'{e}rolles\\
CH-1700 Fribourg, Switzerland}
\date{\today}
\maketitle
\tightenlines
\widetext
\advance\leftskip by 57pt
\advance\rightskip by 57pt

\begin{abstract}
The absorption and emission spectra of transitions between a localized level
and a two-dimensional electron gas, subjected to a weak magnetic field,
are calculated analytically. Adopting the Landau level bosonization
technique developed in previous papers, we find an exact expression for
the relative intensities of spectral lines. Their envelope function,
governed by the interaction between the electron gas and the core hole, 
is reminescent of the famous Fermi edge singularity, which is 
recovered in the limit of a vanishing magnetic field. 
\end{abstract}

\vskip 1cm

]

\narrowtext
\tightenlines

Optical absorption and emission processes from core states in metals are
expected to show striking many-body effects close to the threshold energy.%
\cite{mahan1} This so-called X-ray edge singularity is dominated by two
competing effects. On the one hand by the {\it Anderson} orthogonality
catastrophe,\cite{anderson} leading to a strong reduction of intensity and,
on the other hand, by the attractive electron-core-hole interaction, leading
to the power law divergence at the X-ray absorption edge predicted by {\it %
Mahan}.\cite{mahan2}

A unified description of the two effects was first provided by {\it %
Nozi\`{e}res and de Dominicis},\cite{nozieres} by treating the
electron-core-hole interaction as a one-body scattering potential, switched
on suddenly at the time of the X-ray transition. They found that the nature
of this behavior - divergent or convergent - is controlled by the scattering
phase shifts.

A particularly transparent analysis of the problem was subsequently given by 
{\it Schotte and Schotte} \cite{schotte} in terms of a bosonization scheme,
inspired by previous work on the {\it Tomonaga} model.\cite{tomonaga} On the
one hand, the orthogonality catastrophe is viewed as a consequence of the
infinite number of low energy electron-hole pairs generated in the vicinity
of the Fermi surface (shake-up) due to the sudden appearance of the
scattering potential. On the other hand, the {\it Mahan} effect is
interpreted as a consequence of the interference between the bosons created
in the shake-up and those representing the extra electron in the
bosonization scheme.

The theoretical predictions appear to agree with X-ray spectra for the first
five metallic elements of the periodic table,\cite{mahan3} although the
interpretation of the spectra is rendered difficult by possible band
structure effects. Advances in the growth of modulation-doped semiconductor
heterostructures have made it possible to investigate the same kind of
processes in quasi-two-dimensional electron gases. In a pioneering work, 
{\it Skolnick et al.}\cite{skolnick} found evidence for a Fermi edge
singularity in the low-temperature photoluminescence spectra of InGaAs-InP
quantum wells. This is surprising since in interband transitions the hole
recoil is expected to damp the edge singularity.\cite{gavoret_muller,sham1} 
{\it Skolnick et al.} argued that the minority carriers are localized due to
alloy fluctuations and thus behave like core holes in X-ray emission. This
interpretation agrees with experiments on cleaner samples \cite{nash} where
the effect is considerably weaker.

The experiments were also carried out in the presence of a perpendicular
magnetic field,\cite{skolnick,nash} where the situation is drastically
changed, since the conduction-band states are quantized into discrete Landau
levels. In this case, the sudden appearance of the hole potential generates
excitations, in which electrons are promoted from one Landau level to
another, across the Fermi energy. As a result, both {\it Mahan} and {\it %
Anderson} phenomena show up in the relative intensities of the discrete
emission peaks.\cite{skolnick,nash} A theory of optical and magneto-optical
phenomena in quasi-two-dimensional electron gases was given in a series of
papers by {\it Hawrylak}\cite{pawel} and {\it Uenoyama} and {\it Sham}.\cite
{sham1,sham2}

Inspired by the work of {\it Schotte} and {\it Schotte},\cite{schotte} we
adopt a bosonization scheme to study the magneto-optical spectra of
electronic transitions between a localized non-degenerate level and the
Landau levels of a two-dimensional electron gas. This allows us to derive
for the first time, to the best of our knowledge, an {\it analytical}
expression for the relative intensities of the emission peaks.

Our model is very similar to the one used in the study of core level
absorption spectra in metals\cite{mahan2,nozieres,schotte} and in previous
works on two-dimensional electron systems.\cite{sham1,pawel,sham2} The
Hamiltonian consists of three contributions 
\[
H=H_{0}^{(e)}+H_{0}^{(h)}+H_{I}^{(e-h)}. 
\]

The first term describes independent electrons (taken here as spinless) of
effective mass $m$ in a magnetic field. For a disk geometry and the
symmetric gauge, $H_{0}^{(e)}$ can be diagonalized in terms of angular
momentum eigenstates,\cite{laughlin} 
\[
H_{0}^{(e)}=\sum_{n=0}^{\infty }\sum_{m_{z}=-n}^{N_{\phi }-n}(\hbar \omega
_{c}n-\mu )c_{nm_{z}}^{\dagger }c_{nm_{z}},
\]
where $c_{n,m_{z}}^{\dagger}$ creates an electron in a Landau level $n$
with azimuthal angular momentum $\hbar m_{z}$, $\omega_{c}=\frac{-eB}{mc}$
is the cyclotron frequency, $\mu $ is the chemical potential (including the
zero point energy $\hbar \omega _{c}/2$) and $N_{\phi}=\frac{BS}{\phi_{0}}$
is the number of flux quanta $\phi _{0}$ piercing a sample of area $S$.

The second term describes a localized hole and is written as 
\[
H_{0}^{(h)}=\hbar \omega _{0}d^{\dagger }d, 
\]
where $d^{\dagger }$ creates a hole in a non-degenerate level of energy $%
-\hbar \omega _{0}$ (measured from the chemical potential). Its wave
function is assumed to be strongly localized at the origin and to have
s-wave symmetry.

The third contribution refers to the electron-hole coupling, which we
represent by a spherically symmetric potential. Angular momentum is
therefore conserved and the potential scatters electrons from $(n,m_{z})$ to 
$(n^{\prime },m_{z})$. Following {\it Schotte} and {\it Schotte},\cite
{schotte} we represent this coupling by a contact potential at the origin, 
\[
H_{I}^{(e-h)}=V_{0}\Psi ^{\dagger }(0)\Psi (0)d^{\dagger }d, 
\]
where $\Psi (0)=\sum_{n,m_{z}}\psi _{n,m_{z}}(0)c_{nm_{z}}$is the fermion
field operator at $\vec{r}=0$. This simplifies considerably the analysis,
since wave functions with $m_{z}\neq 0$ vanish at the origin, $\psi
_{n,m_{z}}(0)=\frac{\delta _{m_{z},0}}{\sqrt{2\pi \ell ^{2}}}$ (where $\ell =%
\sqrt{\frac{\hbar }{m\omega _{c}}}$ is the magnetic length), and are
therefore not affected by the localized hole. Then, defining $\delta
_{B}/\pi =-V_{0}\rho _{F}$ with $\rho _{F}=\frac{m}{2\pi \hbar ^{2}}$ being
the density of states for $B=0$, we arrive at the simplified electron-hole
interaction Hamiltonian 
\[
H_{I}^{(e-h)}=-\hbar \omega _{c}\frac{\delta _{B}}{\pi }\sum_{n^{\prime
},n}c_{n^{\prime },0}^{\dagger }c_{n,0}d^{\dagger }d. 
\]
Analogously to the {\it Schotte} and {\it Schotte} paper, $\delta _{B}$ can
be identified as the phase shift in the Born approximation. We remark that
the scheme developed in this paper can be easily generalized to the case of
finite range spherically symmetric potentials.

For simplicity, we restrict ourselves to the case of a relatively weak
magnetic field, tuned in such a way that the $N$ electrons fill completely
the $\nu =N/N_{\phi }$ lowest Landau levels. This corresponds to an integer
quantum Hall regime with large filling factor. Thus the ground state of $%
H_{0}^{(e)}$ is given by 
\begin{equation}
\left| G_{0}\right\rangle =\prod_{n=0}^{\nu -1}\prod_{m_{z}=-n}^{N_{\phi
}-n}c_{n,m_{z}}^{\dagger }\left| 0\right\rangle .  \label{|G>}
\end{equation}

At this point we adopt a Landau level bosonization scheme, introduced in
previous papers,\cite{lbos1} to express all the excitations of the
electronic system in terms of bosonic fields.

In a first step we enlarge the Hilbert space by including negative $n$
states, in the spirit of the {\it Luttinger} model.\cite{luttinger}
Accordingly, the ground state (\ref{|G>}) has to be redefined, so that all
the negative energy states are completely filled. These unphysical states
should have little influence on the low-energy excitations.

Next we introduce the bosonic operators 
\[
b_{n,m_{z}}^{\dagger }=\frac{1}{\sqrt{n}}\sum_{p=-\infty }^{\infty
}c_{n+p,m_{z}}^{\dagger }c_{p,m_{z}}\text{ for }n\geq 1, 
\]
which generate neutral excitations above the ground state. In order to
bosonize the fermion operators, we use the phase representation 
\[
c_{m_{z}}^{\dagger }(\theta )\equiv \frac{1}{\sqrt{2\pi }}%
\sum_{n}e^{in\theta }c_{n,m_{z}}^{\dagger }. 
\]
In close analogy to bosonization in one dimension,\cite{haldane,voit} we
express these operators in terms of bosonic fields, 
\begin{equation}
c_{m_{z}}^{\dagger }(\theta )=\frac{1}{\sqrt{2\pi \varepsilon }}e^{i\nu
\theta }e^{i\Theta _{m_{z}}^{\varepsilon }(\theta )}U_{m_{z}}.  \label{cbos}
\end{equation}
Here $U_{m_{z}}$ is a unitary operator which increases the number of
fermions with angular momentum $m_{z}$ and provides us with the appropriate
anticommutation properties. $\Theta _{m_{z}}^{\varepsilon }(\theta )$,
defined for each angular momentum channel as 
\[
\Theta _{m_{z}}^{\varepsilon }(\theta )=N_{m_{z}}\theta -i\sum_{n=1}^{\infty
}\frac{e^{-\frac{n\varepsilon }{2}}}{\sqrt{n}}\left\{ e^{in\theta
}b_{n,m_{z}}^{\dagger }-e^{-in\theta }b_{n,m_{z}}\right\} . 
\]
is equivalent to the chiral phase operator of 1-D bosonization,\cite{lbos1}
and $N_{m_{z}}=\sum_{p}c_{p,m_{z}}^{\dagger }c_{p,m_{z}}-\left\langle
c_{p,m_{z}}^{\dagger }c_{p,m_{z}}\right\rangle $ is the charge operator.
Instead of taking the limit $\varepsilon \rightarrow 0$, we can interpret $%
1/\varepsilon $ as a bandwidth cutoff eliminating large $n$ contributions.
This cutoff increases as $\nu $ increases, but the final result is cutoff
independent, as in the solution of {\it Schotte} and {\it Schotte}.\cite
{schotte}

The main difference to the bosonization introduced before\cite{lbos1} is
that here a channel is identified by its angular momentum instead of its
guiding center. This choice is indeed more useful in the present case since
all shake-up processes produced by the spherically symmetric potential
conserve angular momentum.

We describe now emission and absorption processes due to transitions between
the localized level and Landau levels. We assume the selection rule $m_{z}=0$%
, which is justified if the hole wave function is spherically symmetric.\cite
{pawel} In the case of absorption the system is initially in the ground
state (\ref{|G>}). Neglecting the $n$-dependence of the dipole matrix
elements, we find a transition rate 
\[
W_{a}\left( \omega \right) \propto \sum_{f}\left| \left\langle f\right|
c_{0}^{\dagger }\left( 0\right) \left| G_{0}\right\rangle \right| ^{2}\delta
\left( \omega -\omega _{0}+\frac{E_{0}-E_{f}}{\hbar }\right) , 
\]
where the summation is performed over all states $\left| f\right\rangle $ of
the electron gas with one electron more than in the ground state $\left|
G_{0}\right\rangle $. The transition rate can be rewritten as 
\begin{equation}
W_{a}\left( \omega \right) \propto \mathop{\rm Re}\int_{0}^{\infty
}dte^{i\omega t}{\cal F}_{a}\left( t\right) ,  \label{A}
\end{equation}
where 
\begin{equation}
{\cal F}_{a}\left( t\right) =e^{-i\omega _{0}t}\left\langle G_{0}\right|
e^{iH_{i}t/\hbar }c_{0}(0)e^{-iH_{f}t/\hbar }c_{0}^{\dagger }(0)\left|
G_{0}\right\rangle .  \label{F(t)}
\end{equation}
The initial and final Hamiltonians $H_{i}$ and $H_{f}$, respectively, can be
represented in bosonized form. Since ${\cal F}_{a}(t)$ involves the
propagation of zero angular momentum states, we can limit ourselves to 
\[
H_{i}=\hbar \omega _{c}\sum_{n=1}^{\infty }nb_{n,0}^{\dagger }b_{n,0}, 
\]
for the Hamiltonian before the absorption (no hole) and 
\[
H_{f}=\hbar \omega _{c}\sum_{n=1}^{\infty }\left\{ nb_{n,0}^{\dagger
}b_{n,0}-\frac{\delta _{B}}{\pi }\sqrt{n}\left( b_{n,0}^{\dagger
}+b_{n,0}\right) \right\} , 
\]
for the Hamiltonian after the absorption (hole present).

Therefore, in terms of the bosons, the hole potential is a simple shift
operator which generates the unitary transformation 
\begin{equation}
T=\exp \left[ \sum_{n=1}^{\infty }\frac{-\delta _{B}}{\pi \sqrt{n}}\left(
b_{n,0}-b_{n,0}^{\dagger }\right) \right] .  \label{T}
\end{equation}

Apart from a constant term that does not affect the dynamics, the
Hamiltonians $H_{f}$ and $H_{i}$ are connected through the relation $%
H_{f}=TH_{i}T^{\dagger }$.

This simplifies the expression (\ref{F(t)}) to 
\[
{\cal F}_a \left( t\right) =e^{-i\omega _{0}t}\left\langle G_{0}\right|
a_{0}(0,t)a_{0}^{\dagger}(0,0)\left| G_{0}\right\rangle , 
\]
where the new fermion operator $a_{m_{z}}(\theta ,t)$ is defined by 
\begin{equation}
a_{m_{z}}(\theta ,t)=e^{iH_{i}t/\hbar } c_{m_{z}}(\theta)Te^{-iH_{i}t/\hbar
}.  \label{am}
\end{equation}

Therefore our task is to evaluate the correlation function of the operator $%
a_{0}(0,t)$ that simultaneously shifts the bosonic fields and creates an
electron in a superposition of Landau levels with zero angular momentum.
This can be easily done by using the bosonized version of the fermion
operators, since the operator $T$ is also an exponential of bosonic fields.
Then, using equations (\ref{cbos}) and (\ref{T}) it is possible to show that 
\begin{eqnarray*}
{\cal F}_a\left( t\right) &=&e^{-i\omega _{0}t}\frac{1}{2\pi \varepsilon }%
\left\langle G_{0}\right| e^{-i\tilde{\Theta}_{0}(0,t)}e^{i\tilde{\Theta}%
_{0}(0,0)}\left| G_{0}\right\rangle \\
&=&\frac{e^{-i\omega _{0}t}}{2\pi \varepsilon }e^{\left(
D_{0}(0,t)-D_{0}(0,0)\right) },
\end{eqnarray*}
where 
\begin{equation}
D_{0}(0,t)=\left\langle G_{0}\right| \tilde{\Theta}_{0}(0,t)\tilde{\Theta}%
_{0}(0,0)\left| G_{0}\right\rangle ,  \label{D}
\end{equation}
and $\tilde{\Theta}_{0}(0,t)$ is written as

\begin{equation}
\tilde{\Theta}_{0}(0,t)={-i}\left( 1-\frac{\delta _{B}}{\pi }\right)
\sum_{n=1}^{\infty }\frac{e^{-n\varepsilon }}{\sqrt{n}}\left\{ e^{in\omega
_{c}t}b_{n,0}^{\dagger }-h.c\right\} .  \label{theta2}
\end{equation}

After substituting the expression (\ref{theta2}) into (\ref{D}) we arrive at 
\[
D_{0}(0,t)-D_{0}(0,0)=(1-\frac{\delta _{B}}{\pi })^{2}\ln \left( \frac{%
\varepsilon }{1-e^{-i\left( \omega _{c}t-i\varepsilon \right) }}\right) , 
\]
that transforms the correlation function (\ref{F(t)}) into 
\[
{\cal F}_{a}(t)=\frac{e^{-i\omega _{0}t}}{2\pi \varepsilon }\left( \frac{%
\varepsilon }{1-e^{-i\left( \omega _{c}t-i\varepsilon \right) }}\right) ^{(1-%
\frac{\delta _{B}}{\pi })^{2}}. 
\]
Inserting this result into equation (\ref{A}) we finally obtain the
absorption spectrum 
\begin{equation}
W_{a}\left( \omega \right) \propto \sum_{n=0}^{\nu -1}\frac{\Gamma \left(
n+\alpha +1\right) }{\Gamma \left( n+1\right) \Gamma \left( \alpha +1\right) 
}\delta ^{\varepsilon }(\omega -\omega _{0}-n\omega _{c})  \label{Ifinal}
\end{equation}
where $\alpha =-2\delta _{B}/\pi +\left( \delta _{B}/\pi \right) ^{2}$, $%
\Gamma (x)$ is the gamma function and $\delta ^{\varepsilon}(x)$ is a
distribution of width $\varepsilon $ that tends to a delta function as $%
\varepsilon \to 0$ .

The calculation of the emission spectrum proceeds in an analogous way.
Initially the system is in the ground state $|G \rangle $ of the full
Hamiltonian $H$ with a hole in the localized level. The transition rate is
given by 
\[
W_e \left( \omega \right) \propto \sum_f \left| \left\langle f \right| c_0
\left(0 \right) \left| G \right\rangle \right|^2 \delta \left(\omega -
\omega_0 - \frac{E_0^{\prime}-E_f}{\hbar} \right), 
\]
where $E_0^{\prime}$ is the ground state energy of $H$, or 
\begin{equation}
W_e \left( \omega \right) \propto \mathop{\rm Re}\int_{0}^{\infty
}dte^{-i\omega t}{\cal F}_e\left( t\right) ,  \label{B}
\end{equation}
where 
\begin{equation}
{\cal F}_e\left( t\right) =e^{i\omega _{0}t}\left\langle G\right|
e^{iH_{i}^{\prime}t/\hbar }c_{0}^{\dagger}(0)e^{-iH_{f}^{\prime}t/\hbar
}c_{0}(0) \left| G\right\rangle .  \label{E(t)}
\end{equation}
The initial Hamiltonian for emission is the final Hamiltonian for absorption
and vice versa, i.e. $H_i^{\prime}=H_f, H_f^{\prime}=H_i$. Furthermore the
ground states of the Hamiltonians before $(|G \rangle )$ and after the
emission $(|G_0 \rangle )$ are related by $(|G \rangle )=T(|G_0 \rangle )$,
where $T$ is the unitary transformation (\ref{T}). This allows us to arrive
at 
\[
{\cal F}_e \left( t\right) =e^{i\omega _{0}t}\left\langle G_{0}\right|
a_{0}^{\dagger}(0,t)a_{0}(0,0)\left| G_{0}\right\rangle , 
\]
where $a_{m_z} \left(\theta ,t \right)$ is again given by equation (\ref{am}%
).

The rest of the calculation proceeds exactly as in the case of absorption
and we find 
\[
{\cal F}_e \left( t\right) = e^{-2i \omega _0t} {\cal F}_a \left( t\right) . 
\]
The emission spectrum $W_e(\omega)$ has the same form as equation (\ref
{Ifinal}), with $n \omega_c$ replaced by $-n \omega_c$.

\begin{figure}
\epsfysize=3.5in
\centerline{\epsfbox{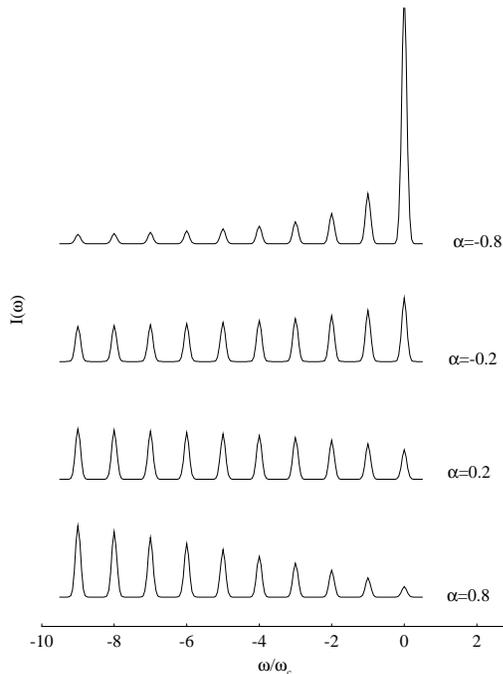}}
\caption{Emission spectra for $\nu=10$ and $\varepsilon=0.1$ with $\omega$ 
measured from the treshold. The envelope of the peaks is reminiscent of 
the $B=0$ case.
\label{fig1}}
\end{figure}

Thus, both the emission and absorption spectra consist of sets of peaks with
intensities proportional to $\frac{\Gamma (n+\alpha +1)}{\Gamma \left(
n+1\right) \Gamma (\alpha +1)}$. This is the central result of this paper.
The emission spectrum is illustrated in Fig.~\ref{fig1}.

In the limit $B\to 0$ (that amounts to $\omega _{c}\to 0$ and $\nu\to\infty$%
), in a small $\omega$ interval there will be a number $n\gg 1$ of
consecutive peaks. Then, by using the asymptotic formula $\frac{\Gamma
(n+\alpha )}{\Gamma (n)}\sim n^{\alpha}$ for $n\gg 1$, we can see that the
envelope of the emission peaks tends to the standard result $I(\omega)\sim
\left( \varepsilon \frac{\omega_0-\omega} {\omega _{c}}\right)^{\alpha}$.
Since physically $\varepsilon^{-1}$ must be of the order of the band width,
which means $\varepsilon \sim \frac{1}{\nu}= B\frac{S}{N\phi_0}$, the
constant $\varepsilon/\omega _{c}\sim \hbar\rho_FS/N$ plays the role of a
field-independent, non-universal cutoff constant.

When $B\neq 0$ there is a single parameter $\alpha $ governing the relative
intensities of the peaks similarly to the zero field case. What occurs here
is that for $\alpha <0$ there is an enhancement of the emission intensity
due to the enhancement of the number of final states in this type of
process. On the other hand, for $\alpha >0$ we observe a reduction on the
emission intensity that can be related to the {\it Anderson} orthogonality
catastrophe. We emphasize, however, that the well-known power-law
singularities of the $B=0$ case are absent when the magnetic field is turned
on. This can be traced back to the appearance of an energy gap, the
cyclotron energy, $\hbar \omega _{c}$, which preempts the infrared
divergence present when $B=0$. Nevertheless, a remnant of this effect can
still be seen in the envelope function of the discrete peaks in Fig.~\ref
{fig1}. This is in agreement with the numerical results of {\it Uenoyama} and%
{\it \ Sham}\cite{sham1} in the case of infinite hole mass and with the
experimental results of {\it Skolnick et al.}.\cite{skolnick}

In conclusion, we have used the bosonization technique for calculating
exactly the emission and absorption spectra for transitions between a
localized level and the discrete Landau levels of a two-dimensional electron
gas. The intensities of the (equidistant) spectral lines follow an envelope
function which is reminiscent of the singular behavior for zero magnetic
field. A single parameter $\alpha $ determines the line
intensities, and this parameter is nothing else than the exponent of the
zero field edge singularity.

We would like to thank J. A. Brum for fruitful conversations. H.W.
acknowledges full support from the Funda\c{c}\~{a}o de Amparo \`{a} Pesquisa
do Estado de S\~{a}o Paulo (FAPESP). A.O.C. is grateful for partial support
from the Conselho Nacional de Desenvolvimento Cien\'{i}fico e
Tecnol\'{o}gico (CNPq). D.B. wishes to thank FAPESP for support during his
visit at UNICAMP, and M. Potemski for useful information about recent
experimental results.

\end{document}